\documentclass[a4paper]{aa}
\usepackage{graphics,times}

\begin{document}

\title{HD\,172481: a super lithium-rich metal-deficient \\ post-AGB binary with a red AGB companion}

\author{
Maarten Reyniers\thanks{Scientific researcher of the Fund for
Scientific Research, Flanders} \and
Hans Van Winckel\thanks{Postdoctoral fellow of the Fund
for Scientific Research, Flanders}}

\offprints{Maarten Reyniers, \email{maarten.reyniers@ster.kuleuven.ac.be}}

\institute{Instituut voor Sterrenkunde, K.U.Leuven, Celestijnenlaan 200B,
B-3000 Leuven, Belgium}

\date{Received 29 June 2000 / Accepted 16 October 2000}

\titlerunning{HD\,172481}
\authorrunning{M. Reyniers \& H. Van Winckel}

\abstract{
We present in this paper a study on the peculiar supergiant HD\,172481. Its spectral
type (F2Ia), high galactic latitude ($b$\,=\,$-$10\fdg37), circumstellar dust, high
radial velocity and moderate metal deficiency ([Fe/H]\,=\,$-$0.55) confirm the
post-AGB character of this object. A detailed chemical analysis shows slight but
real s-process overabundances, however no CNO-enhancement was detected. Furthermore,
the spectral energy distribution and the TiO bands in the red part of the spectrum
reveal a red luminous companion. The luminosity ratio of the hot F type component
and this cool M type companion L$_{F}$/L$_{M}$ is derived for a reddening of
E(B-V)\,=\,0.44 (L$_{F}$/L$_{M}$ $\sim$\,1.8) and indicates that the companion
must also be strongly evolved and probably evolving along the AGB. Neither our
photometric data-set, nor our radial velocity monitoring show evidence for orbital
variability which may indicate that the period is too large for direct binary
interaction. Most interestingly, a strong lithium resonance line is detected, which
yields an abundance of $\log\epsilon({\rm Li})$\,=\,3.6. Several explanations for
this large lithium content are explored.
}

\maketitle

\section{Introduction}

Although the research on the late stages of stellar evolution gained enormous
momentum thanks to the IRAS survey back in the beginning of the eighties, a detailed
understanding of the final stellar evolution in general and of the post-AGB phase in
particular, is still lacking. Since the evolutionary time for an object to cross the
HR diagram
from the tip of the AGB to the planetary nebula phase is short (100-10000 years
depending on the core-mass; e.g. Bl\"ocker 1995) the objects are rare and the
significance of results on individual objects is difficult to extrapolate to the
whole AGB stellar evolution theory.

Moreover, in recent years it became clear that binarity can influence the final
stages in many ways. This complicates the general observational picture since
results on disguised binaries may lead to misinterpretations when confronted with
single star evolution theory. Stars like the Red-Rectangle, which were longtime
considered to be proto-typical post-AGB objects, were later discovered to be evolved
interacting binaries (Waelkens et al. 1996). More recently, ISO results on the
detailed infrared analysis of the circumstellar dust spectra revealed that also the
dust evolution can be quite different in binaries than in single stars (e.g. Waters
et al. 1998, Molster et al. 1999). There is growing observational evidence that in
evolved binaries with a wide period distribution, part of the circumstellar material
is trapped somewhere in the system. This has a strong impact on the chemical and
dynamical evolution of the system (e.g. Van Winckel et al. 1995, 1999; Jura \&
Kahane 1999). Thanks to the trapped dust, binaries will have a prolonged IR
lifetime so any sample of post-AGB objects defined on IR measurements will be biased
towards binarity (Waters et al. 1997; Van Winckel 1999).

Since the stellar luminosity increases by large factors (a typical L$_{*}$ is
between 3000-8000\,L$_{\sun}$) during stellar evolution, most of the systems with one
component in a post-AGB evolutionary phase, are single-lined spectroscopic binaries
(SB1). Double-lined binaries should be rare since they must consist of two, almost
equally luminous and thus almost equally evolved objects. The initial mass of both
stars must have been therefore extremely similar.

One possible such object was found to be AFGL\,4106 (Molster et al. 1999, Van Loon
et al. 1999) but it was recognized to be a massive double evolved system consisting of
a massive F-type post-red-supergiant and an M-type red supergiant. The spectral energy
distribution (SED hereafter) is complex (see Fig. 2 and 5 in Molster et al. 1999)
with the F-type dominating the UV and optical wavelengths while the M-type component
dominates the near-IR. In the far-IR the huge cool dust component is clearly
observed. The luminosity ratio of L$_{F}$/L$_{M}$=1.8 was found with an estimated
total luminosity of L$_{F}$\,=\,1.3\,$\times$\,10$^{5}$ L$_{\sun}$ assuming a distance
of 3.3 kpc (Molster et al. 1999).

In this paper we discuss in detail another such rare double evolved SB2 system:
HD\,172481 (Table \ref{tab:basics}). 
The binary consists of a luminous F-type star, probably in a post-AGB
evolutionary stage, and an M-type luminous companion, probably evolving on the AGB.
The paper is organized as follows: in Sect. 2 we give an overview of the divers
observational data we acquired during the last years. In Sect. 3 we list all the
observational evidence for the binary SB2 nature of the system; we construct and
discuss the spectral energy distribution (SED) and give the spectral indications of
the cool component. In Sect 4 we focus on the photospherical chemical composition of
the F-type component and discuss the high Li content. The overall discussion on the
system is given in Sect. 5 while the summary of the main conclusions is given in
Sect. 6.

\begin{table}\caption{Basic parameters of HD\,172481 (SIMBAD database).}\label{tab:basics}
\begin{center}
\begin{tabular}{rrr}
\hline \multicolumn{3}{c}{HD\,172481} \\ \hline \hline
Coordinates & $\alpha_{2000}$
& 18 41 36.96 \\
            & $\delta_{2000}$ &$-$27 57 01.2\phantom{0} \\
\hline Galactic    & $l$ &    6.72  \\ coordinates & $b$ &$-$10.37  \\ \hline

Mean Magnitude   &   B & 9.58 \\
            &   V & 9.09 \\
\hline
Spectral type &  & F2Ia0 \\
\hline
\end{tabular}
\end{center}
\end{table}

\section{Data}
\subsection{Photometry}
We obtained 54 photometric datapoints of HD\,172481 in the Geneva photometric system
during several runs (1989-1996) within the framework of our photometric monitoring
program of optically bright post-AGB candidate stars (e.g. Bogaert 1994). The Geneva V
magnitudes are displayed Fig. \ref{fig:fotorv}. The object is found to be clearly
variable with a small peak to peak amplitude of 0.22 magn. in V, much
larger than the estimated observational error of 0.005--0.007 magn (Rufener \&
Nicolet 1988). The near-IR JHKLM
photometry was taken with the ESO\,1m telescope during the night of 15 to
16/8/92. For the construction of the SED (Sect. \ref{subsec:sed}) we used the mean
magnitudes of the Geneva and near-IR data complemented with photometry from Humphreys
(1975), Olsen (1984) and the IRAS PSC via the SIMBAD database.
The Geneva photometric measurements can be obtained from the authors upon request.

\subsection{Radial Velocities}
\begin{figure}
\caption{Geneva photometry (upper panel) and radial velocity measurements
(lower panel) of HD\,172481.}\label{fig:fotorv}
\resizebox{\hsize}{!}{\includegraphics{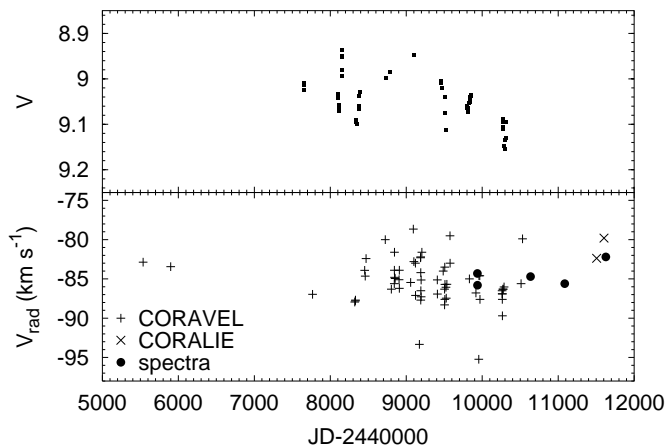}}
\end{figure}

Our sample of radial velocities was obtained with the CORAVEL radial velocity
spectrometer installed at the Danish\,1.5m telescope at La Silla, Chile. Since
the spectrum is correlated with a hardware mask of Arcturus (K2III), always the same
lines contribute to the radial velocity determination which makes this sample very
homogeneous. The Arcturus mask can be used for supergiants with spectral type up
to F0 and the velocities have an accuracy of $\sim$0.5\,km\,s$^{-1}$ (Baranne et al.
1979). Some recent velocity measurements obtained with the CO\-RA\-LIE
spectrograph mounted on the new Swiss Euler telescope at La Silla and also the
velocities from our high resolution spectra (Table \ref{tab:obslog} for details)
were added to this sample. All velocity data are plotted in Fig. \ref{fig:fotorv}.

\subsection{Spectra}\label{subs:spectra}
High resolution, high signal-to-noise spectra were taken with the Utrecht Echelle
Spectrograph (UES) mounted on the 4.2m William Herschel Telescope (WHT) located on La
Palma, Spain, with the EMMI Spectrograph mounted on the 3.5m New Technology
Telescope (NTT) located in La Silla, Chile, and with the Fiber-fed Extended Range
Optical Spectrograph (FEROS) installed at the ESO 1.52-m telescope also at La Silla.
A log of the observations can be found in Table \ref{tab:obslog}.

The WHT and NTT spectra were standard reduced using the specific echelle package of
the MIDAS data analysis system and included order tracing, bias correction, cosmic
hits identification and cleaning, scattered light subtraction and flat fielding. The
orders were extracted by using a simple sum of the pixels and the wavelength
calibration was performed using a ThAr spectrum. During the NTT measurement of 1998
the internal focus of the instrument camera jumped with a small amount of encoder
steps, so the full resolution was not obtained. For the FEROS spectrum, we used the
pipeline reduction which includes all steps of a standard echelle reduction. All
spectral orders were normalized by fitting the continuum through interactively
defined continuum points. Sample spectra can be found in Fig.
\ref{fig:hbeta},
\ref{fig:TiO},
\ref{fig:H_alpha} and
\ref{fig:dubbel}.

\begin{table*}\caption{Log of the observations (spectra), together with an
approximate value for the signal-to-noise of the different spectra, the
heliocentric radial velocities $v_{\rm helio}$ and the number of spectral lines
used in the velocity determination. The accuracy of these velocities is about
1.5\,km\,s$^{-1}$, except for the velocities derived from the EMMI spectra
($\sim$2.5\,km\,s$^{-1}$). The s/n ratio was determined by computing the
standard deviation in a stellar continuum region.}\label{tab:obslog}
\begin{center}
\begin{tabular}{rrrrrrl}
\hline
Date and UT    & Telescope+   & Sp. Range & Resolving &  s/n & s/n & $v_{\rm helio}$ \\
               & Spectrograph &   (nm)    & Power $\lambda/\delta\lambda$ & @\,447\,nm & @\,603\,nm & (km\,s$^{-1}$)   \\
\hline
 6/08/1995 22:34 & WHT+UES  &364-456.5& & 135 & & $-$84.3$\pm$1.7 (56 lines)\\
 7/08/1995 22:33 & WHT+UES  &  453-680& {\raisebox{2.5ex}[0pt]{$\sim$50,000}} & &175 &$-$85.8$\pm$1.6 (130 lines)\\
 6/07/1997 07:46 & NTT+EMMI & 580-1044& & & 140 & $-$84.7$\pm$2.3 (45 lines)\\
30/09/1998 03:06 & NTT+EMMI & 580-1044&
{\raisebox{2.5ex}[0pt]{$\sim$60,000}} & & 85 &$-$85.6$^1$\\
23/03/2000 09:48 & ESO1.5+FEROS &353-921.5&$\sim$48,000& 130 & 120 & $-$82.2$^2$\\
\hline
\multicolumn{7}{l}{\scriptsize $^1$correlation technique with EMMI spectrum of 6/07/1997; $^2$correlation technique with UES spectrum of 7/08/1995}
\end{tabular}
\end{center}
\end{table*}

\section{HD\,172481 as SB2}

\subsection{TiO-bands}

\begin{figure}
\caption{Comparison of the H$_{\beta}$ region of HD\,172481 (WHT+UES) and the
massive F supergiant HR\,1865 confirming the F-type spectrum of HD\,172481.}\label{fig:hbeta}
\resizebox{\hsize}{!}{\rotatebox{-90}{\includegraphics{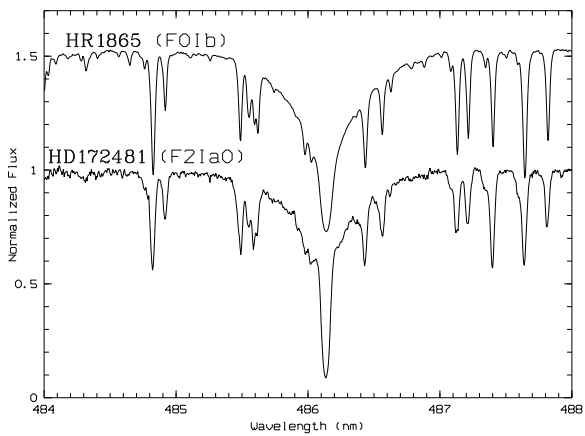}}}
\end{figure}

\begin{figure*}
\caption{ The EMMI and FEROS spectra in the 700-715\,nm TiO band
head region. The spectra are offset by 0.3 units. The TiO bands in the red
spectrum of HD\,172481 are a clear signature of its cool companion. Moreover, these
bands vary in time as evidenced by our 1998 EMMI spectrum. Band head wavelengths are
from Valenti et al. (1998).}\label{fig:TiO}
\resizebox{\hsize}{!}{\rotatebox{-90}{\includegraphics{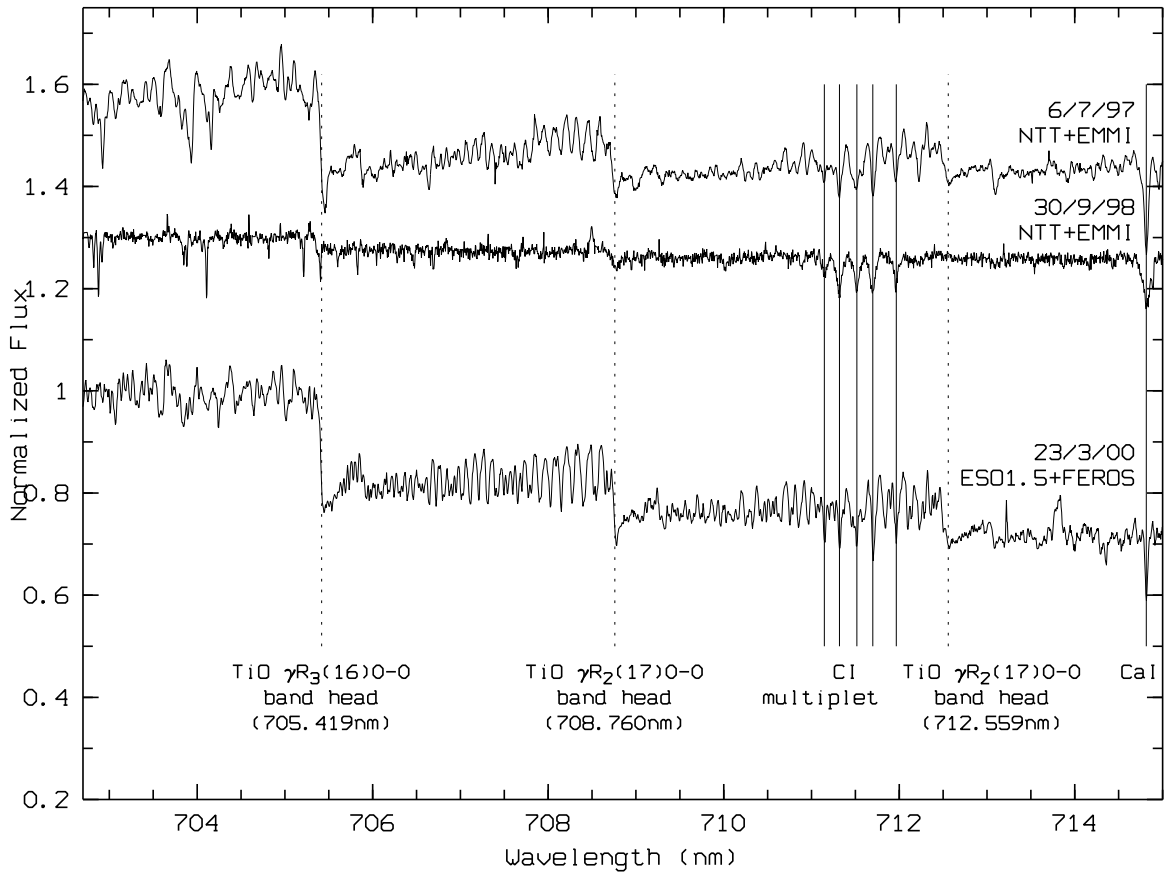}}}
\end{figure*}

The optical high-resolution spectra confirm the F-spectral type given in SIMBAD
which is illustrated in Fig. \ref{fig:hbeta}. However, in the 1995 UES spectra, we
realized a decreasing S/N with increasing wavelength, and we attributed this to the
increasing contribution of a cool companion. Our assumption was confirmed by the
detection of strong red TiO band heads in our 1997 EMMI spectra, and later in the
2000 FEROS spectra (Fig. \ref{fig:TiO}). This is a clear indication that a red
companion is contributing to the total spectrum of HD\,172481. Moreover, the
strength of these bands is clearly variable as evidenced by our 1998 EMMI spectra
(see also Fig. \ref{fig:TiO}), pointing to a high amplitude variation of the
companion. In the latter spectrum, only weak TiO band heads can
be seen, but a clear CI multiplet could be identified. Since the TiO bands are
clearly visible, the cool component must be of spectral type M but a more precise
spectral classification is difficult due to the composite nature of the spectrum.

\subsection{SED}\label{subsec:sed}

\begin{figure}
\caption{The dereddened SED of HD\,172481 with a E(B-V) = 0.44. The dotted line
represents the Kurucz model with T$_{\rm eff}$=7250\,K and log(g)=1.5; the dashed
line the model with T$_{\rm eff}$=3500\,K and log(g)=0.5; the full line the combined
model.}\label{fig:SED}

\resizebox{\hsize}{!}{{\includegraphics{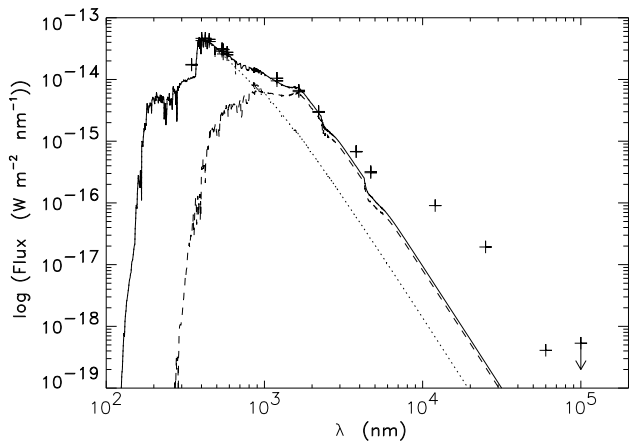}}}
\end{figure}

In Fig. \ref{fig:SED}, the spectral energy distribution (SED) of HD\,172481 is
compiled from the photometric data. We used model atmospheres of Kurucz (Kurucz 1993)
to fit the data. We adopted the atmospheric parameters of our abundance
analysis (Sect. \ref{sec:analysis}; T$_{\rm eff}$\,=\,7250\,K, $\log g$\,=\,1.5). By
calculating synthetic Geneva colour indices using the passbands from Rufener \&
Nicolet (1988) and the above Kurucz model, we obtained a total reddening of
E(B-V)\,=\,0.44, which is somewhat higher than the value published by Bersier (1996):
E(B-V)\,=\,0.271. Using a single object, the fit of only the F-type model atmosphere
clearly did not satisfy from about $\lambda\sim10^3$\,nm redwards and the spectral
characteristics were unlike dust emission. To fit all the data we invoked a second,
cool component, with T$_{\rm eff}$=3500\,K but the model parameters of the cool
component are only derived by a few photometric points and are therefore highly
uncertain. Since, however, the F component is clearly a luminous object, also the M
component must be luminous. The luminosity ratio of both components is found to be
L$_{F-type}$/L$_{M-type}$\,=\,1.8 but this is rather dependent on the adopted
reddening: a lower reddening of E(B-V)\,=\,0.2 would decrease this ratio to 1. The
integration of the dereddened photometry gives a luminosity of the F-type component
of 600 d$^{2}$ L$_{\sun}$ with d in kpc.

In Table~\ref{tab:lamref} we list the contribution of the F-star to the total
composite spectrum for some reference wavelengths. From about 975 nm redwards, the
cool component is more luminous than the F-component.

\begin{table}
\caption{The contribution in percent of the F-type component to the total composite
spectrum as deduced from the SED fit. This information is therefore only
indicative and does not take into account the variability of the M-type
component (see Fig. \ref{fig:TiO}).}\label{tab:lamref}
\begin{center}
\begin{tabular}{llll} \hline
$\lambda$ (nm)  & \%  &  $\lambda$ & \% \\ \hline
301 & 100 & 801 & 60 \\
401 & 99  & 901 & 52 \\
501 & 96  & 1002 & 48 \\
601 & 84  & 1102 & 43 \\
701 & 74 & 1202 & 37 \\ \hline
\end{tabular}
\end{center}
\end{table}

In addition to the two stellar components, the IRAS fluxes ($f_{12} = 5.41 {\rm\,Jy}$,
$f_{25} = 5.22 {\rm\,Jy}$, $f_{60}  = 0.59 {\rm\,Jy}$, $f_{100} < 1.85 {\rm\,Jy}$),
which are also plotted on Fig. \ref{fig:SED}, point to a cool infrared excess which
we interpret as caused by a circumstellar dust shell or disk. A small excess seems to
be already present in the L (3.8 $\mu$m) and M (4.7 $\mu$m) photometric bands
and might indicate a present day mass-loss of the cool component. Clearly more
infrared data is needed to characterize the properties of the M-type component.

We can conclude that the SED consists of 3 components: a luminous F component
dominating the UV and visible part of the spectrum, a luminous M component
dominating the near-IR and a cool dust envelope dominating the IRAS fluxes.

\subsection{Photometry and Radial Velocity}\label{sec:photrad}

\begin{figure}
\caption{Detail of the light curve of HD\,172481. The harmonic fit with a period of
P=91.7d is only considered to indicate the typical timescale of the semiregular
variability of HD\,172481.}\label{fig:fotodet}
\resizebox{\hsize}{!}{\includegraphics{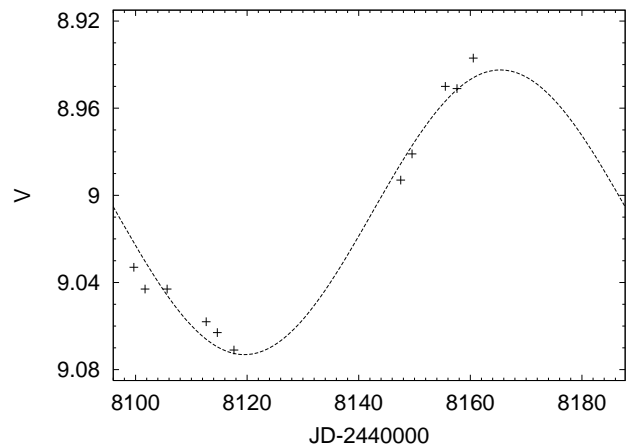}}
\end{figure}

It is clear from Fig. \ref{fig:fotorv} that HD\,172481 is variable, in magnitude as
well as in radial velocity. Despite the poor sampling, several techniques ({\sc
clean} Roberts et al. 1987, {\sc pdm} Stellingwerf 1978, and {\sc scargle} Scargle
1982, Horne \& Baliunas 1986) were used to find a period (pulsational or
orbital) in these data, but none of them yielded a convincing stable period, in the
sense that the folded light curve never satisfied. Nevertheless, by fitting details
of the light curve as has been done in Fig. \ref{fig:fotodet}, we can conclude that
HD\,172481 is a semi-regular pulsator varying with a  timescale of about
$\sim$100 days.

One can argue that the variability in the photometry is induced by a variable
contribution of the M-type companion in the Geneva V filter. However, this seems
very unlikely because the latter filter is centered around 549\,nm where the
contribution of the cool companion is only $\sim$10\% (Table \ref{tab:lamref}).
Moreover, this filter correlates well with the U photometric filter
($\lambda_c$\,=\,346\,nm) for which the contribution is negligable. Even in the G
filter ($\lambda_c$\,=\,581\,nm), no clear trace of a variable contribution could
be found. In addition, one would expect a larger amplitude if the pulsation is
(also) due to the cool AGB star: an amplitude of 0.2 mag. is rather usual for
post-AGB stars.

\begin{figure}
\caption{Normalized spectra of HD\,172481 around H$\alpha$. Each spectrum is offset
by 1 unit.}\label{fig:H_alpha}
\resizebox{\hsize}{!}{\includegraphics{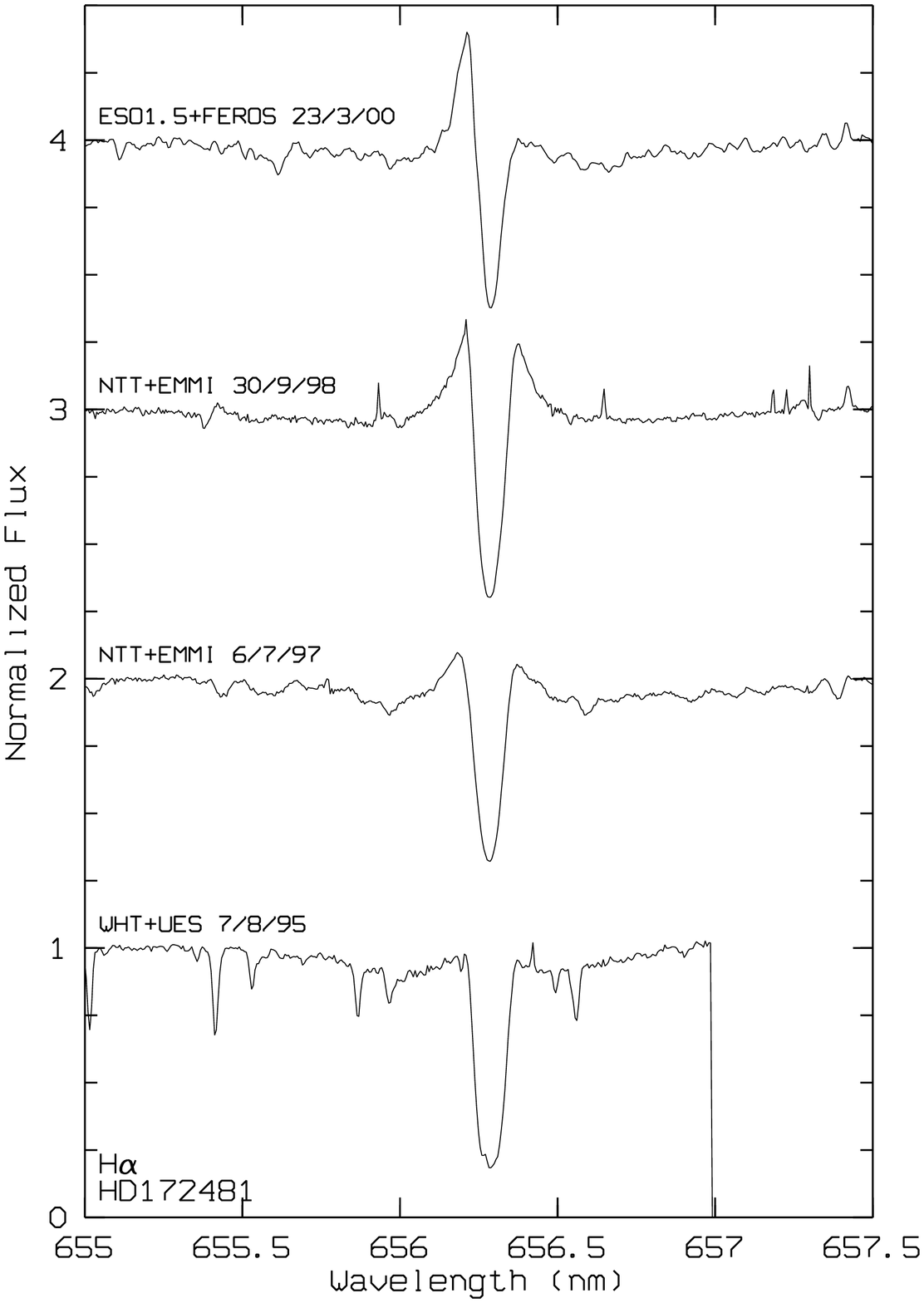}}
\end{figure}

In Fig. \ref{fig:H_alpha} we present our high resolution spectra around the
H$\alpha$ line. The line exhibits a shell type profile, an emission profile
that is found in almost all other optically bright post-AGB candidate stars
(e.g. Waters et al. 1993; Oudmaijer \& Bakker 1994; Van Winckel et al. 1996).
It is still unclear which mechanism is responsible for the variable H$\alpha$
emission in the spectra of post-AGB stars. A recent photometric and
spectroscopic study about the nature of the pulsation of the bright post-AGB
star HD\,56126 (Barth\`es et al. 2000) suggests that the irregular pulsation
probably originates from the interaction between the photospheric dynamics
and the dynamics of the higher atmospheric layers. Shock waves are generated
when the receding photospheric layer collides with the upcoming one and
probably play a major role both in driving the pulsation and limiting its
amplitude.

Although a peak-to-peak amplitude of 17\,km\,s$^{-1}$ is observed, we could not
discover a long-term trend in the radial velocity data which would point to resolved
binary motion. This could be a projection effect, if we view the system on a
unfavourable inclination, but is likely due to the long orbital period and hence
very small velocity amplitude. Assuming typical masses for the post-AGB and AGB
objects (F-type = 0.6 M$_{\sun}$ and M-type = 1.0 M$_{\sun}$), an inclination of
60$^{\circ}$ and a velocity amplitude of 5\,km\,s$^{-1}$, we obtain a period of 51
years. Note that the CORAVEL mask is defined in the blue ($\lambda<520$\,nm)
so the contribution of the M-component in negligable.

We can conclude that both the light curve and the radial velocity data give evidence
for the complex and not well understood variability of the atmospheric layers seen
in most post-AGB stars. This is further strengthened by the variable H$\alpha$
emission. Evidence for binary induced photometric variability or orbital motion is
not observed, which is probably indicative of a long orbital period. The radial
velocity sample indicates too long an orbital period to expect direct interaction
during stellar evolution.

\section{Chemical Analysis}\label{sec:analysis}

We performed a detailed abundance analysis on our UES spectra to gain insight into
the evolutionary status of the F-type component. We confined our analysis to these
two spectra for several reasons: due to the increasing contribution of the cool
companion at longer wavelengths, we decided not to use the EMMI spectra since
they concentrate on the red region of the optical spectrum. Since many useful lines
are situated in the blue and green region a significant chemical study can still be
performed on this limited spectral domain. Moreover, the UES spectra display a very
good signal-to-noise.

\begin{figure*}
\caption{Atomic lines of s-process elements and Li\,I and bands of TiO in one spectrum.
On the left panel, lines of s-process elements can be seen, with the remarkable
detection of a Hf\,II line on our UES spectra. On the right panel, the 671\,nm region
is plotted, with the Li\,I doublet. Clearly, the contribution of the cool companion in
this region is larger for our FEROS spectrum (which is offset by 0.3 units) as
evidenced by the detection of the TiO band head at 671.447\,nm.}\label{fig:dubbel}
\resizebox{\hsize}{!}{\rotatebox{-90}{\includegraphics{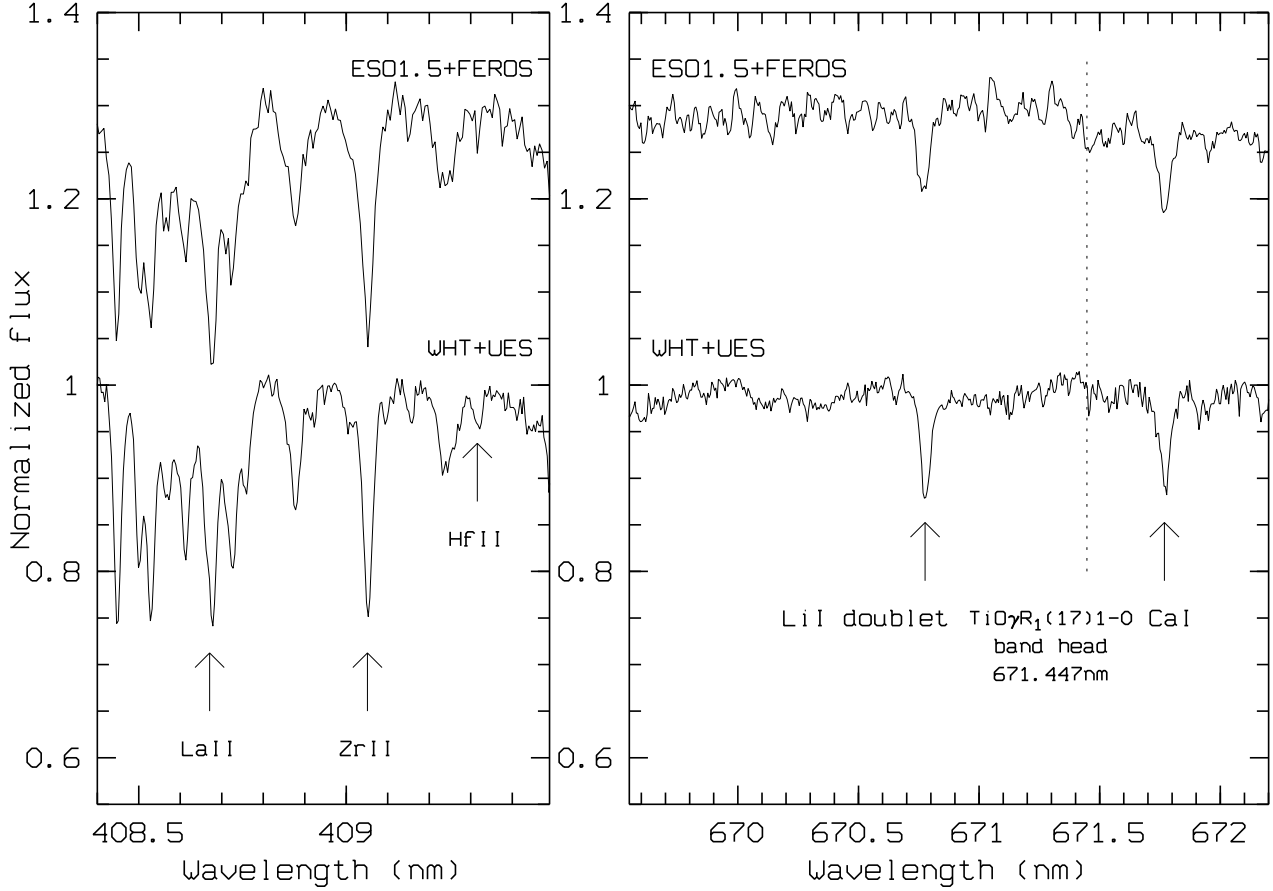}}}
\end{figure*}

Apart from the resolution difference, the quality of the FEROS spectrum is very
similar to the UES spectra (Fig. \ref{fig:dubbel}), but a comparison of the red
region reveals that the contribution of the cool companion was larger during the
FEROS run. Therefore, we preferred to use solely our UES spectra for the chemical
analysis.

\subsection{Line data and Atmospheric Parameters}

In order not to compromise the results of a chemical analysis, one has to
take care that only lines with well determined oscillator strengths ($\log gf$-values)
are selected. A list of such lines useful for the chemical analysis of A and F type
stars has been collected at our institute during the past few years and is still
regularly updated. We described this list in detail in Van Winckel \& Reyniers
(2000). The list we used for HD\,172481 is essentially the same since only a few
lines were added.

We used the CDROM-grid of LTE model atmospheres by Kurucz (1993) in combination with
his abundance calculation programme {\sc width9}. The parameters of an atmospheric
model are determined by forcing the computed abundance of Fe to be independent of
excitation potential (determination of T$_{\rm eff}$), ionisation stage (gravity)
and reduced equivalent width (microturbulent velocity). This method yielded T$_{\rm
eff}$\,=\,7250\,K, log\,$g$\,=\,1.5 and $\xi_t$\,=\,4\,km\,s$^{-1}$. 48 Fe\,I-lines and 14
Fe\,II-lines were used to determine the appropriate atmospheric parameters. With this
model, we also obtain a very good abundance consistency for all other chemical
elements for which lines of two ionisation stages are available (Ca, Cr, Mn and Ni)
confirming an accurate set of atmospheric parameters.

The equivalent widths of 209 lines (28 different ions) used in the analysis were
measured, using direct integration for unblended lines and multiple Gaussian fitting
for blended lines. A rather surprising discovery in the spectra we took is the
presence of the strong Li resonance line at 670.8\,nm (see Fig.~\ref{fig:dubbel} and
\ref{fig:li}).
Before discussing in detail the overall results of the chemical analysis we focus on
the determination of the Li abundance.

\subsection{Li-abundance}
The lithium abundance is derived by spectrum synthesis of the 670.8\,nm region.
Spectrum synthesis is used to account for the doublet structure of the Li
resonance line. Line parameters are taken from Kurucz (1993) and are nearly
the same as in Cunha et al. (1995). Kurucz's program {\sc synthe}
was used to make the syntheses.

\begin{figure}
\caption{The LiI resonance line in the UES-spectrum of HD\,172481. The points are the
observed spectrum, the lines  synthetic spectra with
$\log\epsilon({\rm Li}) = 3.47$, 3.57 and 3.67 resp. A least-squares fit
resulted in $\log\epsilon({\rm Li}) = 3.57$ (full line).}\label{fig:li}
\resizebox{\hsize}{!}{\includegraphics{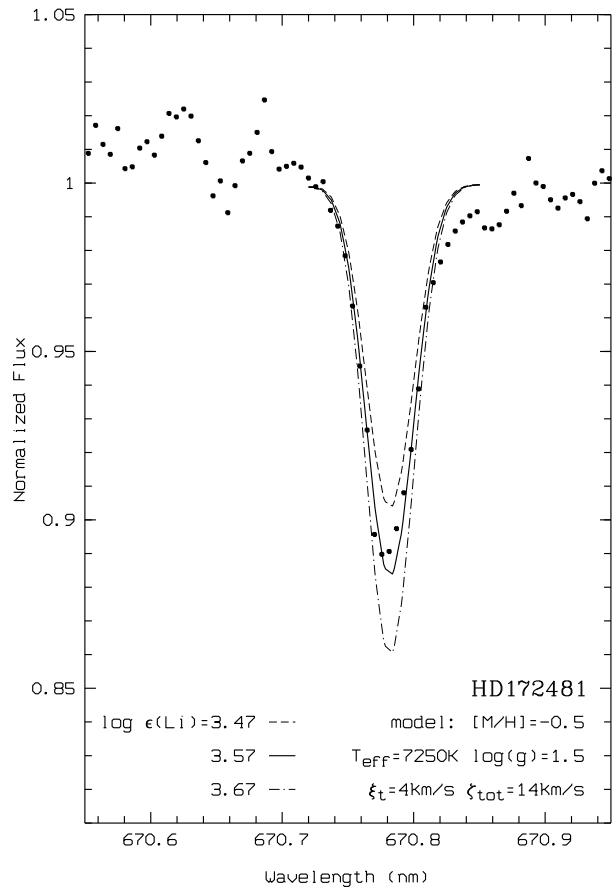}}
\end{figure}

Synthetic spectra have to be convolved with a macroturbulent broadening function and
with the instrumental broadening, to match the line widths of the observed spectrum.
For the instrumental broadening we used a gaussian with a full width at half maximum
given by the spectral resolution of the UES spectrograph:
$\zeta_{\rm instr}$\,=\,6\,km\,s$^{-1}$. The
macroturbulent velocity $\zeta_{\rm macr}$ is determined in the following way.
We selected 17
clearly unblended lines of different width ($<$\,150\,m\AA) and different species.
For each line, we calculated the abundance from the equivalent width using the {\sc
width9}-program. After that, we synthesized with the Kurucz spectral synthesis
programme {\sc synthe} each line using the obtained abundance. The total broadening
velocity $\zeta_{\rm tot}$
($\zeta^2_{\rm tot}$\,=\,$\zeta^2_{\rm instr}$+$\zeta^2_{\rm macr}$)
is the only free parameter in this synthesis. We found a slight
dependence of $\zeta_{\rm tot}$ on the equivalent width of the line:
\[ \zeta_{\rm tot} = 0.09 * W_{\lambda} + 9.18
\] with a classical correlation coefficient of 0.93. We adopted for $\zeta_{\rm tot}$
a value of 14\,km\,s$^{-1}$ (hence $\zeta_{\rm macr}$\,$\simeq$\,12.6\,km\,s$^{-1}$)
for the LiI-line, yielding a Li-abundance of $\log\epsilon({\rm Li}) = 3.57$ using a
least-squares fit (Fig. \ref{fig:li}).

\begin{table}\caption{Sensitivity of the Li-abundance to the stellar
parameters and an upper limit for the non-LTE correction. The standard parameters of
HD\,172481 are T$_{\rm eff}$=7250K, $\log g=1.5$, $\xi_t=4$\,km\,s$^{-1}$ and
$\zeta = 8$\,km\,s$^{-1}$; the standard Li-abundance
$\log\epsilon({\rm Li}) = 3.57$.}\label{tab:sens}
\begin{center}
\begin{tabular}{rrrr}
\hline
parameter             & symbol              & variation & variation \\
                      &                     & &$\log\epsilon({\rm Li})$ \\
\hline
continuum placement   &                     & low/high  & $\pm$0.03\\
temperature &T$_{\rm eff}$& $\pm$250\,K     & $\pm$0.24\\
gravity   & $\log(g)$ & $\pm$0.5 (cgs)      & $\pm$0.11\\
microturb. velocity  & $\xi_t$          & $\pm$2\,km\,s$^{-1}$& $\pm$0.03\\
broadening& $\zeta$   & $\pm$2\,km\,s$^{-1}$& $\pm$0.04\\
\hline
non-LTE correction &  &                     &$<$0.2$^*$\\
\hline
\end{tabular}
\end{center}
{\footnotesize $^* \ \log\epsilon({\rm Li})_{\rm LTE}$$-$$\log\epsilon({\rm Li})_{\rm non-LTE}$,
estimated from Carlsson et al. (1994).}
\end{table}

In Table \ref{tab:sens} we investigate the sensitivity of the derived Li-abundance
to the uncertainties in the stellar parameters. It is clear that the temperature is
by far the most delicate parameter, though a variation of 250K (as in Table
\ref{tab:sens}) is rather large. We estimate the uncertainty of the lithium
abundance around 0.20 dex. To our knowledge, detailed non-LTE calculations of
the Li\,I resonance line for this temperature-gravity domain are not available in
literature, but an extrapolation of the results published by Carlsson et al. (1994)
indicates that a non-LTE correction would not exceed 0.2 dex
(i.e. $0<\log\epsilon({\rm Li})_{\rm LTE}-\log\epsilon({\rm Li})_{\rm non-LTE}<0.2$).

Adopting $\log\epsilon({\rm Li}) = 3.57$, we expect for
the subordinate LiI-line at 610.36\,nm an equivalent width of $W_{\lambda}$\,=\,5\,m\AA.
Unfortunately, this line is blended with a FeII-line ($\lambda$610.350\,nm).
Nevertheless, an upper limit for the Li-abundance can be derived from this line:
$\log\epsilon({\rm Li}) < 3.7$. The LiI-line at 812.645\,nm has a calculated
equivalent width of 0.6\,m\AA\ and is obviously not detectable on our spectra.

\subsection{Overall results of the chemical analysis}

\begin{figure}
\caption{The abundances of HD\,172481 relative to iron [el/Fe].
Errorbars are plotted if the number of lines is more than 3.
The horizontal line represents [el/Fe]=0.
Details of the adopted solar abundances can be found in the caption of
Table \ref{tab:results}.}\label{fig:elfe}
\resizebox{\hsize}{!}{\includegraphics{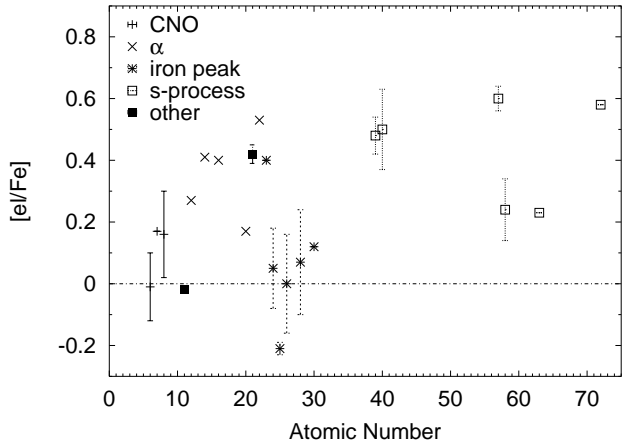}}
\end{figure}

The derived abundances of all other elements are summarized in
Table~\ref{tab:results}; a graphical representation can be found in Fig.
\ref{fig:elfe}. Besides the lithium line at 670.8\,nm, also the resonance line of
europium ($\lambda$\,412.973\,nm) was treated with spectrum synthesis, in order to
account for the strong hyperfine splitting of this line. The hyperfine data were
taken from Biehl (1976).

The {\em carbon} (C) and {\em oxygen} (O) abundances are well established, using the
multiplets at 615.6\,nm and 645.4\,nm for oxygen. Carbon is clearly not enhanced, and
the slight overabundance of oxygen ([O/Fe]=+0.2) is normal in this metallicity range
(e.g. Boesgaard et al. 1999). A {\em nitrogen} (N) abundance was hard to obtain,
because most useful N lines are situated in the red part of the spectrum, which we
did not use in our analysis. The N abundance is derived from very small lines
(equivalent widths 11, 5 and 7 m\AA). Consequently, one has to  interpret the
tabulated abundance ($\log \epsilon({\rm N})=7.61$)  more as an upper limit rather
than as a fixed abundance. We can conclude that also N is not enhanced.

No sodium (Na) enhancement is measured. The {\em $\alpha$-elements} Mg, Si, S, Ca and
Ti are all enhanced with respect to iron, with a simple mean [$\alpha$/Fe]=0.36. Such
a value is expected in this metallicity range (e.g. McWilliam 1997). The titanium
(Ti) abundance is quite high, but it is deduced from rather strong lines (7 out of
the 13 lines have $W_{\lambda}$$>$100\,m\AA) and therefore very sensitive to the
microturbulent velocity: an increase of 1\,km\,s$^{-1}$ in $\xi_t$ decreases the Ti
abundance by 0.1 dex. Scandium (Sc) is also quite high, but this element behaves like
an even-Z $\alpha$-element (Nissen et al. 2000) and it is also known to be slightly
hyperfine structure sensitive. The high vanadium (V) abundance is less easy to
explain, because normally it should follow the iron deficiency (Chen et al. 2000).
The hyperfine structure sensitivity alone cannot explain this overabundance.
We searched for other V\,II lines from the list of Bi\'emont et al. (1989), but,
using the extensive lists of Kurucz (1993) and the VALD-2 database (Kupka et al. 1999),
all turned out to be blended. The {\em iron peak elements} (Cr, Ni, Zn) do follow the
iron deficiency, only manganese (Mn) is slightly underabundant, which is also a
Galactic evolution effect (Nissen et al. 2000).

Moderate but clear {\em s-process} overabundances are found for HD\,172481.
The strontium (Sr) abundance is very difficult to determine accurately since the
resonance lines at $\lambda$421.5\,nm and $\lambda$407.7\,nm are heavily saturated
and weaker optical lines are not present. The other two elements of the Sr-peak
around the magic neutron number 50, yttrium (Y) and zirconium (Zr), have more usable
lines and turn out to be clearly enhanced by 0.5 dex. For the heavy s-process
elements around magic neutron number 82, the situation is less clear, due to the
smaller amount of lines. Nevertheless, we feel confident that also the hs elements
lanthanum (La) and cerium (Ce) are enhanced. Just like Sr, an accurate  barium (Ba)
abundance determination is difficult due to the lack of weak lines. For neodymium (Nd),
we deduced an upper limit of $\log \epsilon({\rm Nd})< 0.9$ ([Nd/Fe]$<$0.0) from the
non-detection of the line at 529.316\,nm. We did the same for samarium (Sm) and
obtained $\log \epsilon({\rm Sm})< 0.55$ ([Sm/Fe]$<$0.1) from the non detection of
the 453.794\,nm line. The r-process element europium (Eu) is slightly overabundant
relative to iron, but this behaviour is expected for this metallicity (e.g. Woolf et
al. 1995). So Nd, Sm and Eu seem {\em not} to be intrinsically enhanced.

Surprisingly, we also detected a small line at $\lambda$409.316\,nm which we
identified as Hf\,II (Fig. \ref{fig:dubbel}). The same line is detected in the heavily
s-process enriched post-AGB stars (Van Winckel \& Reyniers 2000).
An abundance calculation with $W_{\lambda}$=12\,m\AA\ yields a very high [Hf/Fe]=+0.6!
This line is hardly detectable on our FEROS spectrum (also Fig. \ref{fig:dubbel})
and the abundance is extreme but we could not find possible blending lines.
Other possibly detectable Hf lines are all blended. The significance of the high
hafnium abundance should be further investigated.

\begin{table}
\caption{Chemical analysis of HD\,172481.  For every ion we list the number of lines
used N, the mean equivalent width $\overline{W_{\lambda}}$, the absolute abundance
$\log\epsilon$ (i.e. relative to H: $\log\epsilon = \log X/{\rm H} + 12$), the
abundance ratio relative to iron [el/Fe] and the internal scatter $\sigma$, if more
than one line is used. For the solar iron abundance we used the meteoric iron
abundance of 7.51. For the solar C, N and O abundances we adopted resp. 8.57, 7.99
and 8.86 (C: Bi\'emont et al. 1993, N: Hibbert et al. 1991, O: Bi\'emont et al.
1991); the other solar abundances are taken from Grevesse (1989). {\em ss} stands
for spectrum synthesis.}\label{tab:results}
\begin{center}
\begin{tabular}{lrrrrr}
\hline \multicolumn{6}{c}{\rule[-0mm]{0mm}{5mm}{\bf HD\,172481 }}\\
\multicolumn{6}{c}{
$
\begin{array}{r@{\,=\,}l}
{\rm T}_{\rm eff} & 7250\,{\rm K}  \\
\log g  & 1.5\ {\rm(cgs)} \\
\xi_t & 4.0\,{\rm km\,s}^{-1}   \\
{\rule[-3mm]{0mm}{3mm}\rm [Fe/H]} & -0.55\\
\end{array}
$
} \\
\hline
\rule[-3mm]{0mm}{8mm}ion & N  &$\overline{W_{\lambda}}$&$\log\epsilon$ & [el/Fe] &
$\sigma$\\ \hline Li\,I  &1   &{\em ss}& 3.57&      & 0.20 \\ \hline C\,I   & 11 &
36 & 8.01 & $-$0.01 & 0.11 \\ N\,I   &  3 &  8 & 7.61 &   +0.17 & 0.03 \\ O\,I   & 5
& 26 & 8.47 &   +0.16 & 0.14 \\ \hline Na\,I  &  2 & 32 & 5.76 & $-$0.02 & 0.12 \\
Mg\,I  &  1 & 37 & 7.30 &   +0.27 &      \\ Si\,I  &  5 & 16 & 7.41 &   +0.41 & 0.12
\\ S\,I   &  4 & 32 & 7.06 &   +0.40 & 0.09 \\ Ca\,I  & 15 & 55 & 5.99 &   +0.18 &
0.14 \\ Ca\,II &  2 & 30 & 5.92 &   +0.11 & 0.02 \\ Sc\,II &  4 & 85 & 2.97 &
+0.42 & 0.03 \\ Ti\,II & 13 &104 & 4.97 &   +0.53 & 0.15 \\  V\,II &  1 &
 61 &  3.85 &    +0.40 &     \\ 
Cr\,I  &  8 & 21 & 5.17 &   +0.05 & 0.11 \\ Cr\,II & 18 & 61 &
5.17 &   +0.05 & 0.13 \\ Mn\,I  &  3 & 21 & 4.62 & $-$0.22 & 0.01 \\ Mn\,II &  1 &
13 & 4.66 & $-$0.18 &      \\ Fe\,I  & 48 & 62 & 6.92 & $-$0.04 & 0.16 \\ Fe\,II &
14 & 68 & 6.96 &         & 0.15 \\ Ni\,I  & 18 & 27 & 5.77 &   +0.07 & 0.16 \\
Ni\,II &  3 & 47 & 5.78 &   +0.08 & 0.19 \\ Zn\,I  &  2 & 32 & 4.17 &   +0.12 & 0.03
\\ \hline Y\,II  &  6 & 42 & 2.17 &   +0.48 & 0.06 \\ Zr\,II &  9 & 50 & 2.55 &
+0.50 & 0.13 \\ La\,II &  4 & 70 & 1.27 &   +0.60 & 0.04 \\ Ce\,II &  6 & 21 & 1.24
&   +0.24 & 0.10 \\ Nd\,II &    &    & $<$0.9\phantom{0} & $<$+0.0\phantom{0} & \\
Sm\,II &    &    & $<$0.55           & $<$+0.1\phantom{0} & \\ Eu\,II &1&{\em
ss}&0.19 &   +0.23 &      \\ Hf\,II &  1 & 12 & 0.91 &   +0.58 &      \\ \hline
\end{tabular}
\end{center}
\end{table}

\section{Discussion}
\subsection{Post-AGB nature}
In this section we list the arguments for the proposed post-AGB character
of our object.
\begin{enumerate}
\item The high galactic latitude (Tab. \ref{tab:basics}; $b$\,=\,$-$10\fdg37),
\item the large radial velocities (Fig. \ref{fig:fotorv}) and
\item the moderate metal deficiency ([Fe/H]=$-$0.55)
\end{enumerate}
point to a population II membership. Furthermore,
\begin{enumerate}
\setcounter{enumi}{3}
\item the SED (Fig. \ref{fig:SED}) shows the presence of dust,
as a result of previous mass loss probably during the AGB phase.
\item The photometry and the H$\alpha$ profiles show a variability very
similar to other post-AGB stars (Sect. \ref{sec:photrad}).
\item Finally, the slight s-process overabundances are probably the result of
the third dredge-up, when He burning products are brought to the surface.
\end{enumerate}
The last argument is, however, not straightforward. If we expect the signature
of a 3rd dredge-up, also the CNO elements should be enhanced, which is clearly
{\em not} detected: the total CNO abundance is equal to the initial value.

Binary post-AGB stars are often depleted, with a dust disk as an ideal site
for the gas-dust separation (see Van Winckel 1999 for a review). Depletion alters
the abundances profoundly. An ideal element to trace for possible depletion is
zinc (Zn) with its low condensation temperature. A high [Zn/Fe] ratio points to
efficient depletion. For HD\,172481, [Zn/Fe]=+0.12, suggesting that no
depletion process has taken place.

\subsection{Li-rich evolved stars}

Plez et al. (1993) published an abundance analysis of seven luminous AGB stars in
the Small Magellanic Cloud. These moderately metal deficient stars ([Fe/H]=$-$0.5)
are Li rich, C poor and show a low $^{12}$C/$^{13}$C ratio. The chemical pattern
observed in these stars is explained by envelope burning, which is the activation of
the H burning CN cycle at the base of the convective envelope (called ``hot bottom
burning'' HBB). The Li abundance in these stars is typically $\log \epsilon({\rm
Li})=3.0$, which is much larger than expected from standard AGB evolution where Li
has been destroyed and/or diluted due to convection. The mechanism thought to be
responsible for this Li production is the $^7$Be-transport mechanism (Cameron \&
Fowler 1971) in which $^3$He is converted to $^7$Li via the reaction chain
$^3$He($^4$He,$\gamma$)$^7$Be($e,\nu)^7$Li. Theoretical models suggest that this
mechanism can only be efficient in massive AGB stars (4-7 M$_{\odot}$; e.g. Sackmann
\& Boothroyd 1992, Mazzitelli et al. 1999). Also massive {\em galactic} AGB stars
with signs of envelope burning were found (Garc\'{\i}a-Lario et al. 1999), but not
yet extensively chemically analysed.

On the other hand, Li is found in many evolved stars in the AGB or post-AGB
evolutionary phase for which there is strong observational evidence that many of
these stars are low-mass objects with M$<$2$-$3\,M$_{\odot}$: galactic low-mass C
stars have been found with Li excesses of $\log \epsilon({\rm Li})>1$ (Abia et al.
1993, Abia \& Isern 1996, 1997, 2000); unexplained high Li abundances
were also found in some metal poor population II post-AGB stars, which showed strong
evidence of an efficient 3rd dredge-up: IRAS\,22272+5435 (Za\v{c}s et al. 1995),
IRAS\,05341+0852 (Reddy et al. 1997), IRAS\,Z02229+6208 and IRAS\,07430+1115 (Reddy
et al. 1999). It has been suggested by Reddy et al. (1999) that these stars could be
the successors of the Li-rich carbon stars studied by Abia (see references above).
However, a detailed abundance analysis of 12 J-type Li-enhanced carbon stars
(Abia \& Isern 2000) showed no s-process enhancements in these stars and a solar
metallicity while the $^{12}$C/$^{13}$C ratio is low. These J-type carbon stars are
clearly chemically different from the 4 Li-enhanced metal-poor post-AGB stars with
high $^{12}$C/$^{13}$C ratios and strong s-process overabundances.

It is generally known that also among red giants ascending the 1st giant branch
high Li-abundances can be found. Wasserburg et al. (1995) developed a non-standard
mixing mechanism for these RGB stars to explain their low $^{12}$C/$^{13}$C ratio
(dubbed ``cool bottom processing'' CBP)
and Sackmann \& Boothroyd (1999) showed that CBP can account for the high lithium
abundances observed in the RGB Li-rich K giants (de la Reza et al. 1997 and
references therein). Moreover, they suggested that CBP might also occur in the
TP-AGB phase (Boothroyd \& Sackmann 1999). de La Reza et al. (1997) suggested that
{\em all} K giants become Li-rich for a short time. During this period the giants are
associated with an expanding thin circumstellar shell supposedly triggered by an
abrupt internal mixing mechanism resulting in a surface Li enrichment. However,
Jasniewicz et al. (1999) did not find any correlation between Li abundance and
infrared excess in a sample of 29 late-type RGB giants with infrared excess. There
are even Li-rich giants with no circumstellar dust.

Alternative external scenarios to explain the high lithium abundances invoke
the accretion of a substellar object (objects).
Such scenarios gained interest by the discovery of several extra solar
planets and brown dwarf candidates around solar-type stars. Planets could
be even commonly present around PNe (Soker 1999). Engulfment scenarios were 
developed (e.g. Siess \& Livio 1999a,b), but cannot account for Li abundances
exceeding the initial one and therefore fail to explain the very Li rich stars with
$\log \epsilon({\rm Li})\ge2.8$ (Siess \& Livio 1999b). In other scenarios
(Denissenkov \& Weiss 2000) planet engulfment acts as the activating mechanism for
the CBP, producing also these higher Li abundances.

We can conclude that the Li-enhancements in the low-mass C stars and the 4
metal-poor post-AGB giants emphasize our poor knowledge of the dredge-up and mixing
processes or possible accretion during stellar evolution.

\subsection{Li-enhancement in HD\,172481}

It is clear that planet engulfment (alone) cannot explain a Li abundance of
$\log\epsilon({\rm Li})$\,=\,3.57.
``Hot bottom burning'' (HBB) is thought to be responsible for the Li enhancements in
moderately metal-poor luminous AGB stars (estimated initial mass $\sim$7\,M$_{\sun}$)
in the Small Magellanic Cloud (Plez et al. 1993).  This scenario is theoretically
thought to occur only in luminous AGB stars of initial mass 3-7\,M$_{\sun}$ (Sackmann
\& Boothroyd 1992). The low metallicity and the kinematics of HD\,172481 suggest an
old and hence lower mass nature. Another argument against the HBB scenario for
HD\,172481 is the absence of a nitrogen enhancement. During the HBB, C is converted
into N by the CN cycle. But possibly lithium production may have occurred without
burning significant amounts of carbon.

On the other hand, the CBP mechanism may be relevant for the Li enhancement in
HD\,172481. The presence of circumstellar dust is in favour of de La Reza's
scenario, if applicable to low mass AGB stars. Unfortunately, we are not able to
derive a $^{12}$C/$^{13}$C ratio, which could shed some light on the deep mixing
involved.

The recent surprising results of the chemical content of J-type carbon stars (Abia
\& Isern 2000) are interesting in many respects. The O-rich circumstellar dust which
is observed in about 5\%-10\% of these objects (Lloyd Evans 1991) is difficult to
reconcile with standard AGB evolution and led Little-Marenin (1986) and Lloyd Evans
(1990, 1991) to conclude that the material expelled by the M-type progenitor was
trapped and stored in a binary system. Recent CO measurements of J-type carbon stars
confirm this scenario and point to a long-lived reservoir of oxygen rich dust
(Jura \& Kahane 1999). In the scenario proposed by Yamamura et al. (2000) to
explain the ISO-SWS spectra of some J-type carbon stars, these stars are wide
binaries in which the oxygen rich dust was stored around the unseen companion. Note,
however, that no direct evidence of the binary nature of these stars exists yet.
Since no large amplitude radial velocity variations are observed (Barnbaum et al.
1991), the eventual orbits must be rather wide and direct binary interaction by
Roche-Lobe overflow is unlikely to have happened during evolution on the AGB. Whether
the special chemical composition of these stars is linked to the proposed binarity in
wide systems is still an open question.

For HD\,172481 there is direct observational evidence for binarity in a probably
wide system. Also this object shows very peculiar abundances with a high Li
abundance, only very small overabundances of s-process elements and no strong C
enhancement. Here, again, the eventual role of the binary nature of HD\,172481 with
respect to the remarkable chemical composition remains an open question.

\section{Summary}
We presented a study of the peculiar supergiant HD\,172481, revealing it to be
a double lined spectroscopic binary consisting of an F-type post-AGB star and an M
type, probably AGB, companion.
The post-AGB nature of the F type component is deduced from its spectral
type; high galactic latitude; circumstellar dust; high radial velocity;
moderate metal deficiency and slight overabundances of the s-process elements
Y, Zr, La and Ce. The photometry
and the H$\alpha$ profiles show a variability very similar to other post-AGB
stars. The spectral energy distribution and the TiO bands in the red part of
the spectrum reveal binarity with a luminous M-type companion with a luminosity
ratio of L$_F$/L$_M$=1.8 assuming a total reddening of E(B-V)\,=\,0.44. The lack
of a resolved long-term trend in the radial velocity data suggests a wide orbit.
Most interesting, we found a very high lithium abundance of
$\log \epsilon({\rm Li})$\,=\,3.57.
This abundance is difficult to reconcile with a ``hot bottom burning'' scenario
because of the probably low mass nature of HD\,172481. An alternative
explanation can be ``cool bottom processing'' (Sackmann \& Boothroyd 1999).

\begin{acknowledgements}
Both authors acknowledge support from the Fund for Scientific Research -- Flanders
(Belgium). This research has made use of the SIMBAD database, operated at CDS,
Strasbourg, France, and the Vienna Atomic Line Database (VALD2).  The anonymous
referee is warmly thanked for the constructive comments.
\end{acknowledgements}


\end{document}